\documentclass[aps,amsmath,amssymb,floatfix,twocolumn,showpacs]{revtex4-1}

\usepackage{graphicx}
\usepackage{dcolumn}
\usepackage{bm}
\usepackage{amsmath}
\usepackage{amssymb}
\usepackage{color}
\usepackage{bbold}
\usepackage{bm}
\usepackage{hyperref}

\providecommand{\doi}[1]{%
  \begingroup
    \let\bibinfo\@secondoftwo
    \urlstyle{rm}%
    \href{#1}%{%
  \endgroup
}

\newcommand{\cl}[1]{\hat{\mathcal{#1}}}

\newcommand{\ave}[1]{\langle #1 \rangle}

\newcommand{\tr}{\text{Tr\,}}
\newcommand{\comm}[1]{{\color{black}{#1}}}

\begin{document}
\title{Pumping approximately integrable systems}
\author{Florian Lange}
\author{Zala Lenar\v{c}i\v{c}}
\email{zala.lenarcic@thp.uni-koeln.de}
\author{Achim Rosch}
\affiliation{Institute for Theoretical Physics, University of Cologne, Z\"ulpicher Stra{\ss}e 77a, D-50937 Cologne, Germany}

\begin{abstract}
Weak perturbations can drive an interacting many-particle system far from its initial equilibrium state if one is able to pump into degrees of freedom approximately protected by conservation laws. This concept has for example been used to realize Bose-Einstein condensates of photons, magnons, and excitons. Integrable quantum systems, like the one-dimensional Heisenberg model, are characterized by an infinite set of conservation laws. Here we develop a theory of weakly driven integrable systems and show that pumping can induce large spin or heat currents even in the presence of integrability breaking perturbations, since it activates local and quasi-local approximate conserved quantities. The resulting steady state is qualitatively captured by a truncated generalized Gibbs ensemble with Lagrange parameters that depend on the structure but not on the overall amplitude of perturbations nor the initial state. We suggest to use spin-chain materials  driven by terahertz radiation to realize integrability-based spin and heat pumps.
\end{abstract}

%\pacs{
%02.30.Ik, %integrable models
%05.70.Ln, %nonequilibrium and irreversible thermodynamics
%%02.50.Ga, %Markov processes
%05.60.Gg, %quantum transport
%75.10.Pq, %spin chain models
%}

\maketitle

A simple classical example for a weakly driven system is a well-insulated 
greenhouse. Due to the approximate conservation of the energy within
the greenhouse, even weak sunlight can lead to high temperatures in its interior, which can be computed from the simple rate equation for the energy transfer. Similarly, large spin accumulation can be achieved in systems with approximate spin conservation \cite{kikkawa98}. Using approximate conservation of the number of photons, magnons, or exciton polaritons, one can use pumping by light to reach densities which allow for the realization of Bose-Einstein condensates \cite{klaers10,demokritov06,kasprzak06}. Number-conserving collisions induce a quasi-equilibrium state in these systems, which can be efficiently described by introducing a chemical potential whose value is determined by balancing pumping  and decay processes. Related theoretical approaches that describe electron-phonon systems far from equilibrium are so-called two-temperature models \cite{allen87}: here one uses that the energy of the electrons and phonons are approximately separately conserved to introduce two different temperatures for the subsystems.

Integrable many-particle systems, like the one-dimensional (1D) fermionic Hubbard model or the XXZ Heisenberg model, are described by an infinite number of (local or quasi-local) conservation laws \cite{faddeev95,grabowski95,prosen13,mierzejewski15,ilievski15,ilievski16}. In closed integrable systems those prevent the equilibration into a simple thermal state, e.g., after a sudden change of parameters. Instead the system can be described by a generalized Gibbs ensemble (GGE) \cite{rigol07,pozsgay13,fagotti13,wouters14,pozsgay14,ilievski15a,vidmar16,essler16,ilievski16a,deluca16a}
\begin{equation}\label{EqGGE}
\rho_\text{0}\sim\exp\!\left(-\sum_i \lambda_i C_i \right)
\end{equation} 
where $C_i$ are the conserved quantities and $\lambda_i$ the corresponding Lagrange parameters. It has also been shown experimentally \cite{langen15} that GGEs for a Lieb-Liniger model can provide highly accurate descriptions of interacting bosons in 1D. 

Many materials are described with high accuracy by integrable models \cite{mourigal13}, however, weak integrability breaking terms and the coupling to thermal phonons imply that in equilibrium these systems are described by simple thermal states, $\rho_0 \sim e^{-\beta H}$, instead of GGEs. The proximity to the integrable point and the presence of approximate conservation laws leads to enhanced spin or heat conductivities (within linear-response theory) \cite{jung06,jung07,jung07a} and also to a slow relaxation  after a quantum quench (via GGE-prethermalization) towards the equilibrium state \cite{bertini15}. 

\begin{figure}
\center \includegraphics[width=\linewidth]{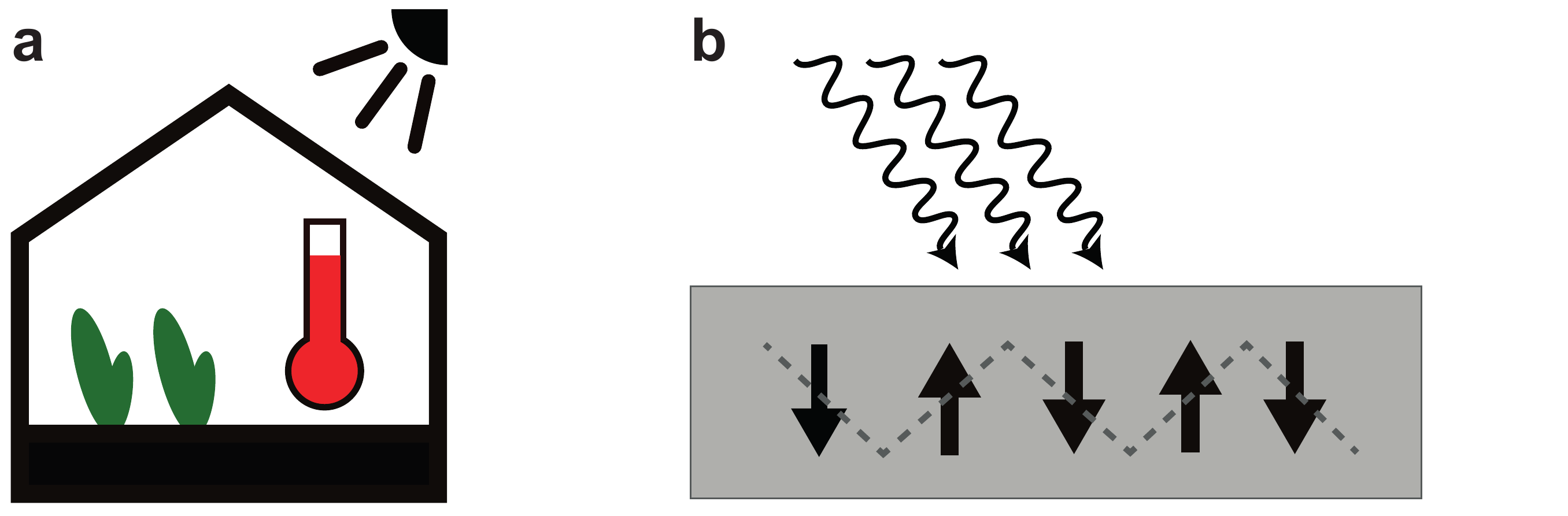}
\caption{(a) \comm{A well-insulated greenhouse exposed to sunshine can heat up significantly since energy within it is approximately conserved. (b) As the heat current in spin chain materials is approximately conserved even weak terahertz radiation can induce large heat current. Material candidates must have appropriate crystal structure, schematically denoted by dashed lines indicating alternating chemical bonds}.\label{fig1}}
%We consider a spin chain material driven by terahertz radiation. Dashed lines indicate schematically alternating chemical bonds. As the heat current is approximately conserved even weak driving can induce large heat currents. \label{fig1}}
\end{figure}

We will show that -- as in the greenhouse example, see Fig.~\ref{fig1} -- such an approximately integrable system can be driven far from its thermal equilibrium by weak perturbations arising, e.g., from a driving periodic in time or from coupling to a non-thermal bath. 
\comm{In order to balance the constant heating due to driving the system has to be
weakly open, e.g., by coupling to a phonon bath. As we will demonstrate this mechanism can be used for example to create large spin and heat currents.}
Besides the quasi 1D systems considered by us, also approximately many-body localized systems are characterized by infinitely many approximate conservation laws which may lead to
a strong response to driving \cite{deLuca15,deLuca16}.

\vspace{.3cm}
\noindent{\bf \large Results}\\
\noindent{\bf Weakly driven system.} 
We consider an interacting many-body system that is approximately described by Hamiltonian $H_0$ and characterized by a finite or infinite number of (quasi-)local conserved quantities $C_i$, $[H_0,C_i]=0$, one of them being $H_0$. 
Energy and other conservations are weakly broken by coupling to thermal or non-thermal baths and/or perturbations periodic in time. For simplicity we assume \comm{periodic boundary conditions} and a (discreet) translational invariance.
We describe the system with density matrix $\rho$ whose dynamics is governed by the Liouvillian super-operator $\cl{L}$,
\begin{equation}\label{heisenberg}
\dot{\rho}=\cl{L} \rho, \quad \cl{L}=\cl{L}_0 + \epsilon \cl{L}_1,	
\end{equation}  
where $\cl{L}$ can be split into the dominant unitary Hamiltonian evolution $\cl{L}_0\rho=-i[H_{0},\rho]$ and perturbation $\cl{L}_1$ of strength $\epsilon$. 
We are interested in the limit of small $\epsilon$ for $t\to \infty$ where a unique (Floquet) steady state $\rho_\infty$ is obtained. 
The general structure of perturbation theory in this case has, e.g., been discussed in Refs.~\cite{cirac92,benatti11,li14}. In this limit, $\rho_\infty$ can be approximated by $\rho_0 = \lim_{\epsilon \to 0} \lim_{t \to \infty} \rho$ with
 $\cl{L}_0 \rho_0=0$ according to Eq.~(\ref{heisenberg}). 
We assume and later support numerically that $\rho_0$ \comm{is approximately described by a GGE}, see Eq.~(\ref{EqGGE}). 

Here it is essential to note that -- as in the greenhouse example discussed above -- the parameters $\lambda_i$ are not
determined by the initial state but by the form of the weak perturbations $\cl{L}_1$.
Our central goal is to compute the $\lambda_i$. We first discuss the case of Lindblad dynamics, where perturbation theory linear in $\epsilon$ can be used, and then focus on Hamiltonian dynamics where we have to consider $\epsilon^2$ contributions. 

\vspace{.3cm}
\noindent{\bf Markovian perturbation.} 
Within the Markovian approximation one can use the
 Lindblad form for $\cl{L}_1$ \cite{petruccione02}. \comm{Note that Lindblad dynamics is considered here mainly for pedagogical purposes (formulas are simpler) while no Lindblad approximation is used for the models studied below.}
The coefficients $\lambda_i$ that fix the GGE are determined from the condition that the change of the approximately conserved quantities 
has to vanish in the steady state
\begin{equation}\label{EqCond1}
\ave{\dot{C}_i}
=\tr(C_i \cl{L} \rho_0) 
= \tr(C_i \epsilon\cl{L}_1 \rho_0)\stackrel{!}{=}0,
\end{equation}
where we used that $\cl{L}_0\rho_0=-i[H_0,\rho_0]=0$. 
Relation (\ref{EqCond1}) yields a set of coupled equations for $\lambda_i$, where the number of equations is equal to the number of conserved quantities. 
We define the super-projector $\hat{P}$ onto the tangential space of GGE density matrix,  
\begin{equation}\label{EqP}
\hat{P} X \equiv -\sum_{i,i'} \frac{\partial \rho_0}{\partial \lambda_i} \ (\chi^{-1})_{ii'} \tr(C_{i'} X),
\end{equation}
using  $\chi_{ii'}=-\tr(C_i \partial\rho_0/\partial\lambda_{i'})$. 
Then the conditions for $\rho_0$ can be compactly written as
\begin{equation}\label{EqCond1a}
\cl{L}_0 \rho_0 = 0, \quad 
\hat{P}(\cl{L}_1 \rho_0)=0.	
\end{equation}
This equation can also be derived by considering higher order perturbations in $\epsilon$, see Methods for details.

\vspace{.3cm}
\noindent{\bf Hamiltonian \comm{perturbation}.} For Hamiltonian dynamics $\epsilon\cl{L}_1 \rho=-i [H_1,\rho]$,
\comm{where $H_1$  may be a sum of several integrability breaking perturbations.}
Perturbation theory linear in $\epsilon$ vanishes, $\tr(C_i \epsilon \cl{L}_1 \rho_0)=0$
for all $\lambda_i$. Therefore one has to expand to order $\epsilon^2$ and Eq.~(\ref{EqCond1a}) is replaced by \begin{equation}\label{EqCond1b}
\cl{L}_0 \rho_0 = 0, \quad 
\hat{P}( \cl{L}_1 \cl{L}_0^{-1} \cl{L}_1 \rho_0)=0.	
\end{equation}
\comm{Since $\hat{P}(\cl{L}_1\rho_0)=0$, $\cl{L}_1 \rho_0$ is not in the kernel of $\cl{L}_0^{-1}$.}
For periodic driving this equation has to be interpreted within the Floquet formalism, see Methods.

\vspace{.3cm}
\noindent{\bf Model.} As discussed in the introduction, our goal is to describe a situation which can be realized experimentally in spin-chain materials driven by lasers operating in the terahertz regime.  
We assume that spin chains are approximately described by a spin-1/2 XXZ Heisenberg model, possibly in the presence of an external magnetic field $B$,
\begin{equation}\label{EqXXZ}
H_{\text{0}}
=\sum_j \frac{J}{2}(S_j^+ S_{j+1}^- + S_j^- S_{j+1}^+) + \Delta S_j^z S_{j+1}^z - B S^z_j.
\end{equation}
The system is driven out of equilibrium by a weak (integrability-breaking) time-dependent perturbation 

\begin{align}\label{EqHd}
&H_\text{d}=\epsilon_\text{d} J \sum_{j}\Big((-1)^{j+1} \mathbf{S}_j\cdot\mathbf{S}_{j+1} \sin(\omega t)+ (-1)^j S^z_j \cos(\omega t) \Big) ,
\end{align}
with driving frequency $\omega$. 
\comm{This specific term has been chosen because it can induce heat and spin currents (as can be shown by a symmetry analysis), and because it can be realized experimentally.}
Such staggered exchange couplings and staggered magnetic fields arise naturally in certain compounds with (at least) two magnetic atoms per unit cell
when coupled to uniform electric and magnetic fields, respectively \cite{affleck99,nojiri06,kimura13,niesen14}. See Fig.~\ref{fig1} for a schematic drawing of such a compound and Methods for concrete experimental suggestions. Therefore $H_{\rm d}$ can be realized by shining a laser (typically at terahertz frequencies) onto the sample. In this case $\epsilon_\text{d}^2$ is proportional to the laser power. Note that for $T=0$ and $B=0$ in the adiabatic limit, $\omega\to 0$, Eqs.~(\ref{EqXXZ},\ref{EqHd}) realize  an adiabatic Thouless pump, where per pumping cycle one spin is transported by one unit cell \cite{shindou05}. We will be interested in the opposite regime of large $\omega$ and large (effective) temperatures.

%Formally the periodic perturbation $H_\text{d}$ would drive the system to infinite temperature \cite{genske15,dalessio14,lazarides14,ponte15} (up to remaining conservation laws \cite{lazarides14a}).
%\commm{The non-interacting models can be an exception \cite{russomanno15}, however already weak time-modulated interactions lead to the trivial steady state via a prethermal-like regime \cite{canovi16}.}
Formally the periodic perturbation $H_\text{d}$ would drive the system to infinite temperature \cite{genske15,dalessio14,lazarides14,ponte15} (up to remaining conservation laws \cite{lazarides14a},
\comm{possibly through a prethermal-like regime \cite{canovi16}}).
In a solid state experiment this is prohibited by the coupling to phonons and, ultimately, to the thermal environment of the experimental setup. We mimic this effect by coupling the spin system to a bath of Einstein phonons, 
$H_0^{\text{ph}}=\omega_{\text{ph}}\textstyle\sum_{j}a_j^\dagger a_j + \dots$, where dots stand for the couplings to further reservoirs which guarantee that the phonon system is kept at fixed temperature $T_{\text{ph}}$, $\rho_{\text{ph}} \sim e^{-H_0^{\text{ph}}/T_{\text{ph}}}$.
\comm{See Methods for details on finite size calculation using a broadened distribution of phonon energies.}
The (weak) coupling to the spin system is described by

\begin{align}\label{EqHph}
H_{\text{ph}}=\epsilon_{\text{ph}} J \sum_{j} &\Bigl(\mathbf{S}_j\cdot\mathbf{S}_{j+1}  (a_j+a_{j}^\dagger) \\
+ &\gamma_\text{m}(S^{x}_j S^{z}_{j+1} + S^{z}_j S^{x}_{j+1}) (a_j+a_{j}^\dagger)\Bigr). \nonumber
\end{align}
To obtain a unique steady state it is essential to break all symmetries, including the $S^z$ conservation. Relativistic effects which relax $S^z$ are mimicked by $\gamma_\text{m}$ in our approach. We expect $\gamma_\text{m} \ll 1$ in materials without heavy elements. For simplicity we set $\gamma_\text{m}=1$ within our numerics as this is found to minimize finite size effects,
\comm{without a qualitative influence on the results.} 
\comm{Besides phonons also other integrability breaking perturbations exist in real materials, including defects, which typically dominate at the lowest temperatures. For high temperatures of the order of $J$ (relevant for the considered setup) it is realistic to assume that phonon coupling dominates. }

In the presence of a periodic perturbation, Eq.~(\ref{EqHd}), in the long-time limit the density matrix is changing periodically, $\rho(t\to \infty)=\sum_n e^{-i \omega n t} \rho^{(n)}$ with ${\rho^{(n)}}^\dagger=\rho^{(-n)}$, $n \in \mathbb{Z}$. 
Within the Floquet formalism one therefore promotes the steady-state density matrix to a vector and Liouville operator to a matrix, see Methods.
For weak driving, $\epsilon_\text{d}\to 0$, only the $n=0$ sector remains and the GGE ansatz, Eq.~(\ref{EqGGE}), simply reads $\rho_0^{(n)}=(\rho_0  \otimes \rho_{\rm ph})\,\delta_{n,0}$ where we included also the phonon density matrix, see above.

\vspace{.3cm}
\noindent{\bf Steady state.} We will use two different approaches to determine  an approximate solution for the steady state density matrix. First, we will parametrize $\rho_0$, Eq.~(\ref{EqGGE}), with a small number of (quasi-)local conserved quantities, $C_i$, $i=1,\dots,N_\text{C}$. In an alternative approach, feasible for small systems, we take all conserved quantities into account: local and non-local, commuting and non-commuting. 
While the second approach is formally exact in the limit $\epsilon_\text{d}, \epsilon_{\text{ph}} \to 0$, the first one is, perhaps, more intuitive and can be computed for larger system sizes.

For the XXZ Heisenberg model an infinite set of mutually commuting local conserved quantities $C_i$ is known, see Methods. $C_1$ is the total spin $C_1=\sum_i S^z_i$ and $C_2=H_{\text{XXZ}}$. Importantly, $C_3$ is the heat current \cite{zotos97}, $C_3=J_\text{H}(B=0)$.
In addition there also exist (infinite) sets of quasi-local commuting conserved quantities \cite{prosen13,mierzejewski15,ilievski15}. As shown in \cite{prosen11,prosen13} the spin-reversal parity-odd family has an overlap with the spin current $J_\text{S}$ at $\Delta<J$. Therefore both heat and spin current could show a large response to a weak perturbation. 
\comm{For our analysis we choose three or five $(N_\text{C}=4,N_\text{C}=6)$ most local conserved quantities $C_i $, $i=1,...,N_\text{C}-1$.}
From the quasi-local sets we include as a single (effective) operator the conserved part of spin current $J_\text{S}^\text{c}$, computed numerically \cite{jung07,mierzejewski14}. For details see Methods. In the presence of an external magnetic field, Eq.~(\ref{EqXXZ}), the heat current also has, in addition to $C_3$, a spin current component, $J_\text{H}=C_3 - B J_\text{S}$.

\begin{figure}
\center \includegraphics[width=\linewidth]{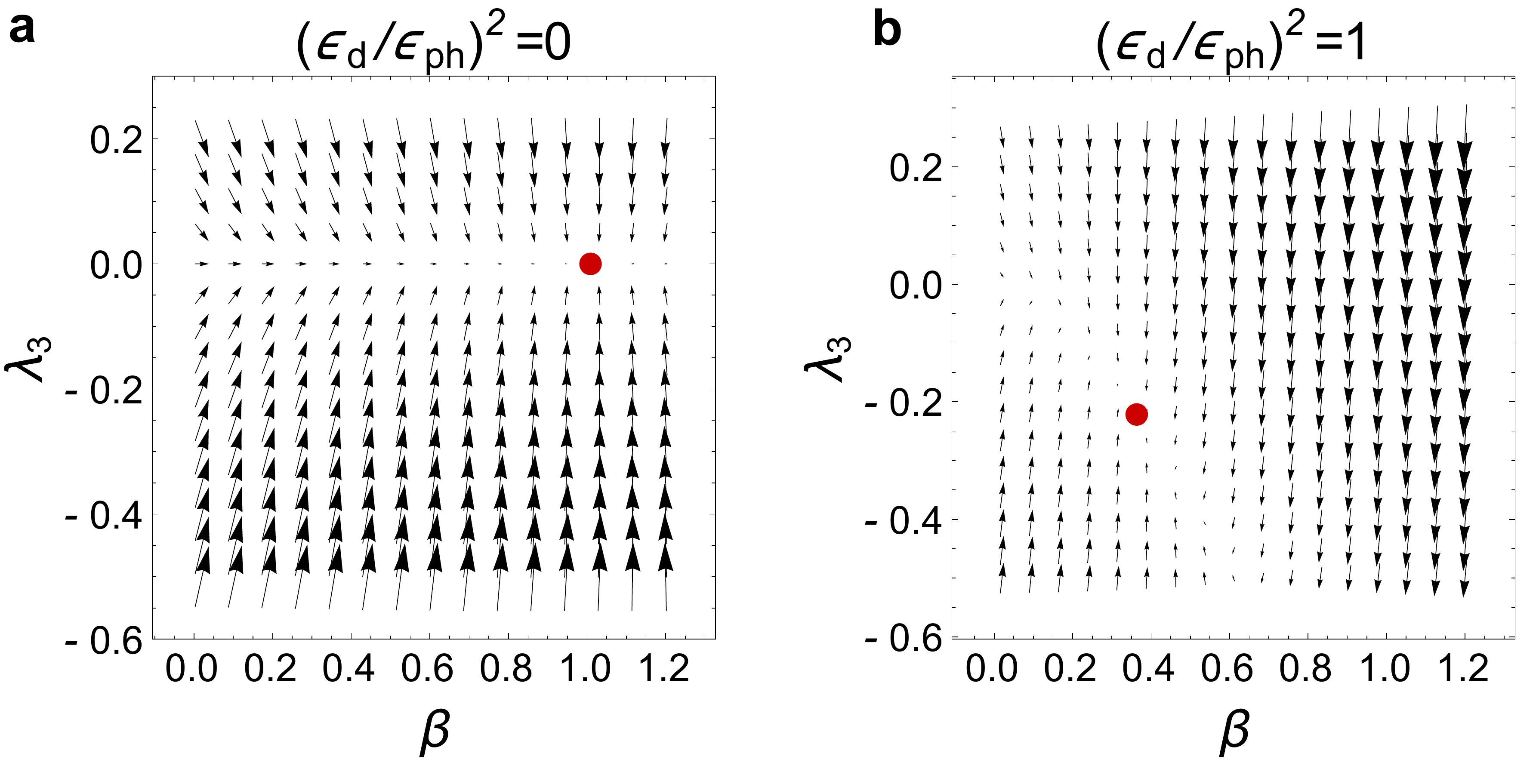}
\caption{Effective force $\boldsymbol{F}$ in the space of Lagrange parameters $(\beta, \lambda_3)$ using $e^{-\beta H_0 -\lambda_3 C_3}$ as an ansatz for the generalized Gibbs ensemble. 
Parameters: $J=\Delta=-B=\omega=\omega_{\text{ph}}=T_{\text{ph}}$. Lagrange parameters $(\beta, \lambda_3)$ are plotted in units $1/J$ and $1/J^2$, respectively. (a) In the absence of an external driving, $\epsilon_\text{d}=0$, the stable fixed point (red dot) is given by the thermal ensemble, $\beta=1/T_{\text{ph}}$, $\lambda_3=0$. (b) When the system is driven by $H_\text{d}$ ($\epsilon_{\text{d}}=\epsilon_{\text{ph}}$), it heats up and $\lambda_3$ becomes finite. \label{fig2}}
\end{figure}
For the visualization of our results it is useful to define generalized forces $F_i$ in the space of Lagrange parameters by rewriting
$\hat{P} \dot \rho=\sum_i \frac{\partial \rho_0}{\partial \lambda_i} F_i$ such that $\dot \lambda_i \approx F_i$, 
\begin{equation}\label{EqLambdadynamics}
F_i=\sum_{i'}
(\chi^{-1})_{ii'}  \tr(  C_{i'} \ \epsilon\cl{L}_1 \cl{L}_0^{-1} \epsilon\cl{L}_1 \rho_0)
\end{equation}
computed using exact diagonalization, see Methods. 
The vector $\bf F$ is a function of the Lagrange parameters $\lambda_i$ which points into the direction of the steady state stable fixed point obtained from $F_i=0$. 
In the absence of driving (Fig.~\ref{fig2}a) one obtains the expected thermal state with $T=T_{\text{ph}}$ while all other Lagrange parameters $\lambda_i$ vanish. For finite driving the GGE is activated and the $\lambda_i$ become finite  (Fig.~\ref{fig2}b). To obtain the steady state, we solve $\chi {\bf F}=0$ using Newton's method.

For the second approach, performed on small $N$-site systems, we first numerically construct a basis in the set of all \comm{(local and non-local)} conserved operators,
$\mathcal{Q}=\{ |n\rangle \langle m| \text{ with } E^0_m=E^0_n \}$, where $H_0|n\rangle=E_n^0 |n\rangle$. Due to degeneracies we find (for finite $B$ and $\Delta \neq J$) about $2 \cdot 2^N$ elements $Q_i \in \mathcal{Q}$. 
In the limit $\epsilon_\text{d}, \epsilon_{\text{ph}}\to 0$ the steady state density matrix $\rho_\infty$ has to fulfill $\cl{L}_0 \rho_\infty=0$ and therefore can be exactly written as a linear combination of the $Q_i$, $\rho_\infty=\sum \alpha_i Q_i$. Using  Eq.~(\ref{EqCond1b}), we therefore find that 
the steady state density matrix for $\epsilon_\text{d}, \epsilon_{\text{ph}}\to 0$ is exactly given by the unique eigenvector with eigenvalue zero of the matrix
\begin{align}\label{EqRhoExact}
&\mathcal{L}^\mathcal{Q}_{mn}=-\tr(Q_m^\dagger \ \epsilon\cl{L}_1 \cl{L}_{0}^{-1} \epsilon\cl{L}_1  Q_n),
\end{align}
where $\cl{L}_0, \cl{L}_1$ are Floquet matrices, see Methods.
\comm{Note that only the relative $\epsilon_\text{d}/\epsilon_{\text{ph}}$ and not the absolute strength of perturbations determine $\rho_0$, as can be seen by dividing the 
equations $\chi \boldsymbol{F}=0$ or $\mathcal{L}^\mathcal{Q}\rho_0=0$ by $\epsilon_{\text{ph}}^2$.}

\begin{figure}[!b]
\center 
\includegraphics[width=.92\linewidth]{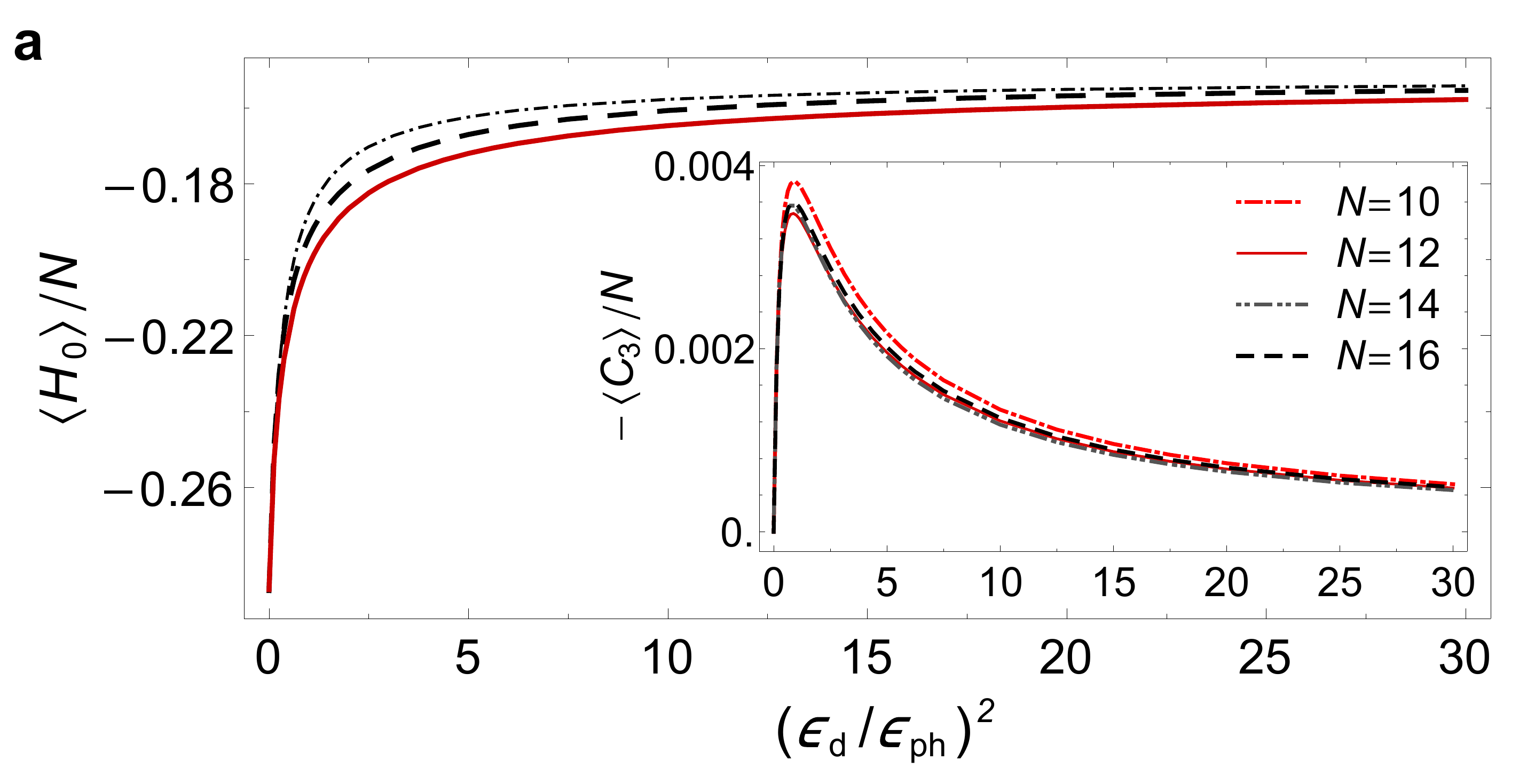}
\includegraphics[width=.9\linewidth]{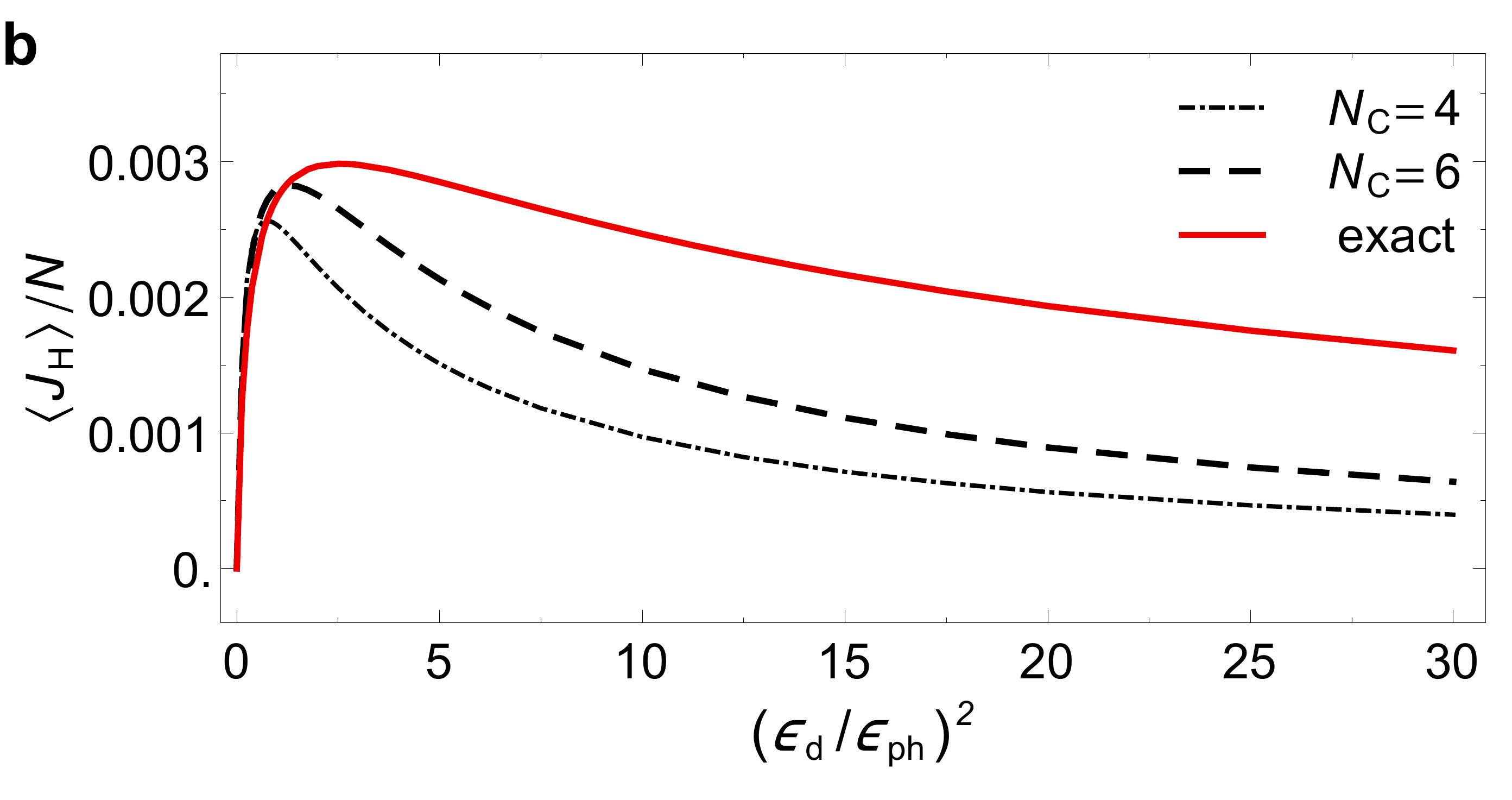}
\caption{Expectation values of (a) energy and (b) heat current densities for a weakly driven spin chain, $\epsilon_{\rm d}, \epsilon_{\rm ph} \to 0$, as functions of the ratio of driving strength $\epsilon_\text{d}$ and phonon coupling $\epsilon_{\rm ph}$. Red solid lines: exact result taking into account all $7969$ conservation laws of a system of $N=12$ sites.
(a) For the energy accurate results are already obtained with a GGE ensemble based on $N_\text{C}=4$ (dot-dashed lines) or $N_\text{C}=6$ (dashed lines) conserved quantities. (b) Also the heat current $J_\text{H}=C_3-B J_\text{S}$ is qualitatively well described by the GGE ensemble but quantitative deviations are larger. Inset: \comm{Finite size analysis for (local) $C_3$ based on GGE ensemble with $N_\text{C}=6$ conserved quantities.}
Parameters: $J=1,\Delta=0.8,B=-1.0,\omega=1.6 \ \omega_{\text{ph}}, \omega_{\text{ph}}=T_{\text{ph}}=1$. \label{fig3}}
\end{figure}

\begin{figure}[!b]
\includegraphics[width=.91\linewidth]{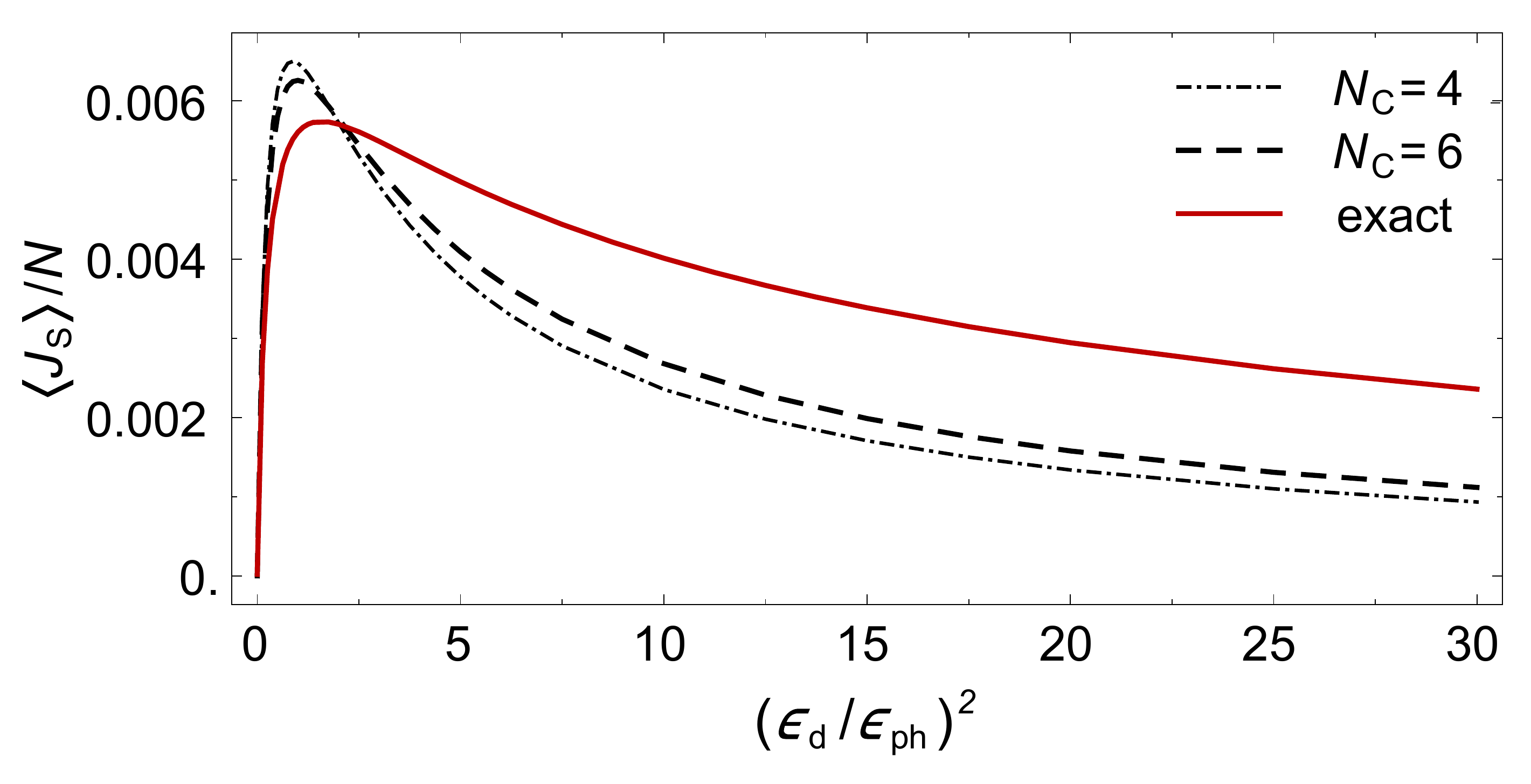}
\caption{For vanishing magnetic field a spin current (but no heat current) is generated within our model for finite ratios of $\epsilon_\text{d}/\epsilon_{\text{ph}}$. The expectation value of spin current density is again maximal for $\epsilon_\text{d}/\epsilon_{\text{ph}}\approx 1$. 
Parameters: $J=1,\Delta=0.8,\omega=1.6 \ \omega_{\text{ph}}, \omega_{\text{ph}}=T_{\text{ph}}=1, N=12$.
 \label{fig4}}
\end{figure}

In Fig.~\ref{fig3} we show the expectation value of the energy and of the heat current densities as functions of $\epsilon_{\rm d}/\epsilon_{\rm ph}$ taking into account $N_\text{C}=4$, $N_\text{C}= 6$, and all conserved quantities. The energy density expectation value is already obtained with good accuracy for $N_\text{C}=4$ and even better for
$N_\text{C}=6$. The  heat current vanishes both in thermal equilibrium, $\epsilon_\text{d}\to 0$, and for $\epsilon_{\text{ph}} \to 0$, where the system is described by an infinite temperature state with finite magnetization, $\rho_0 \sim e^{-\lambda_1 S^z}$ and $\langle H_0 \rangle =-B \langle S_z \rangle$. It takes its largest value for
$\epsilon_{\rm ph} \sim \epsilon_{\rm d}$. 
\comm{For the currents a description in terms of $N_{\rm C}=4$ or $6$ is qualitatively but not quantitatively accurate. Our study strongly suggests that further quasi-local conserved quantities contribute, as discussed in quench protocols \cite{wouters14,pozsgay14,ilievski15a}, see also Ref. \cite{jung07}.}
For the chosen parameters our results depend only weakly on the system size $N$, see inset of Fig.~\ref{fig3}.
\comm{System size analysis is performed for $N_{\rm C}=6$ since the solution based on all conservations cannot be obtained for larger systems.}

\begin{figure}[!b]
\center \includegraphics[width=.9\linewidth]{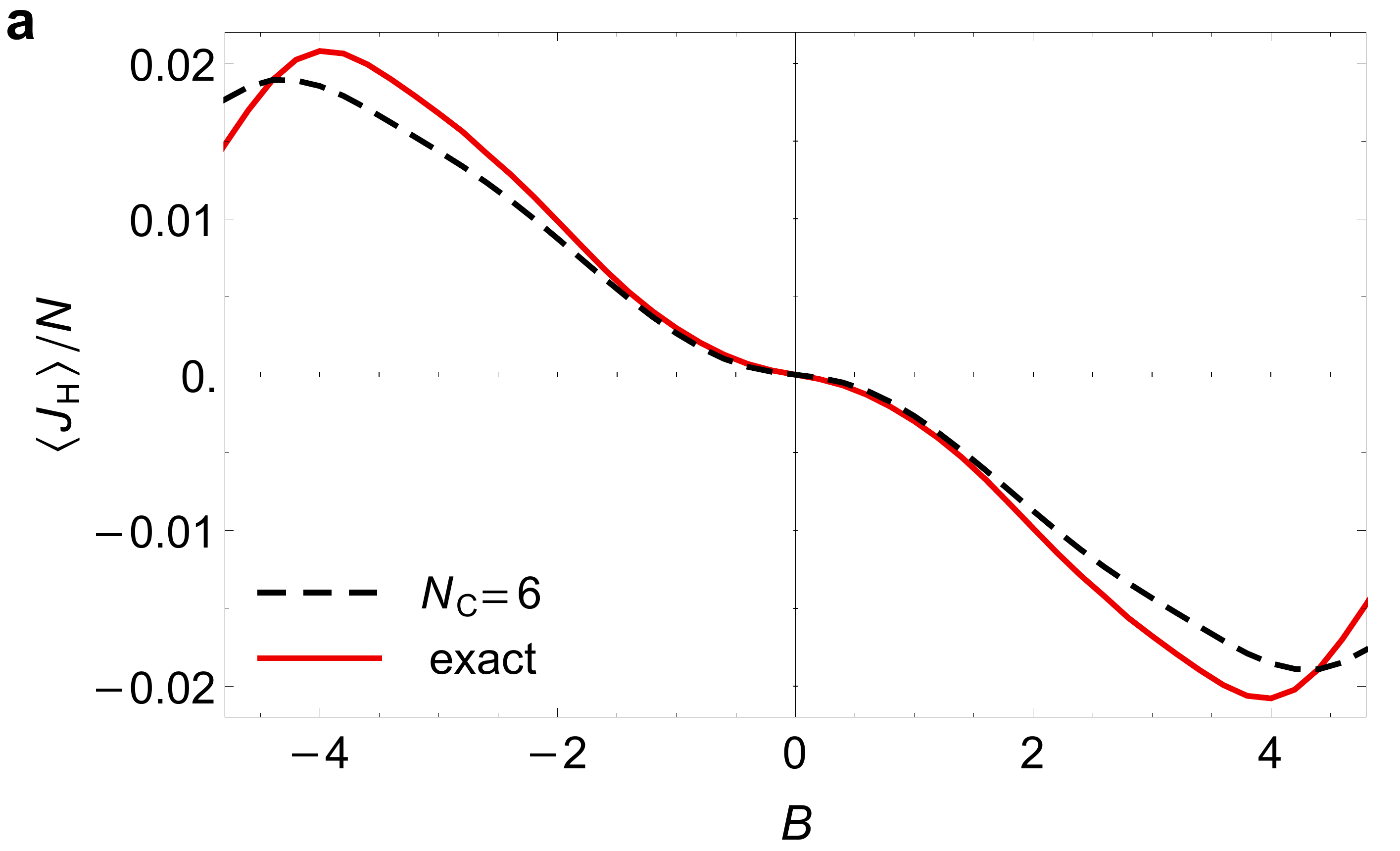}\\
\vspace{-0.2cm}
\center \includegraphics[width=.9\linewidth]{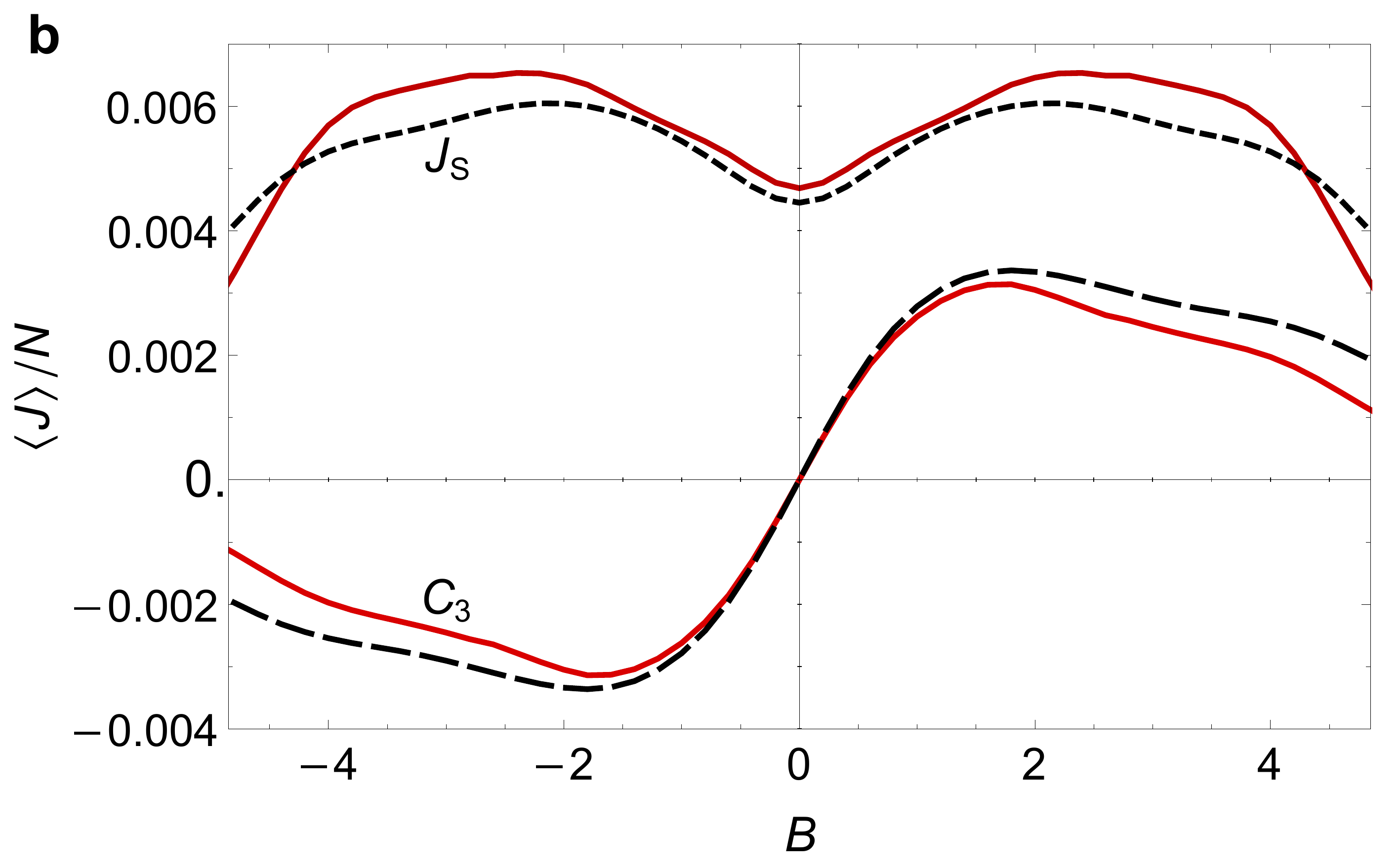}
\caption{
(a) Heat current $J_{\rm H}$, (b) spin current $J_{\rm S}$, and $C_3$ densities as a function of external magnetic field $B$ obtained from a GGE ensemble with $N_{\rm C}=6$ conserved quantities (dashed) or from an exact calculation (solid) including all conservations.
Parameters: $(\epsilon_{\rm d}/\epsilon_{\rm ph})^2=2.5$, $J=1,\Delta=0.8,\omega=1.6 \ \omega_{\rm ph}, \omega_{\rm ph}=T_{\rm ph}=1, N=12$.
 \label{fig5}}
\end{figure}

Our setup can also be used to create spin currents.  
Whilst, by symmetry (bond-centered rotation in real and spin space by $\pi$ around $y$ axis), a finite external field $B$ is needed to obtain a finite heat current, this is not the case for the spin current.  
Fig.~\ref{fig4} displays the spin current density as a function of $\epsilon_{\rm d}/\epsilon_{\rm ph}$ for $B=0$. Qualitatively one obtains a behavior rather similar to the results for the heat current shown in  Fig.~\ref{fig3} with a maximum in the spin current for $\epsilon_{\rm ph} \sim \epsilon_{\rm d}$.

The external magnetic field $B$ is a parameter which can easily be tuned experimentally. Fig.~\ref{fig5} shows heat and spin current densities as a function of external magnetic field $B$ for $(\epsilon_{\rm d}/\epsilon_{\rm ph})^2=2.5$. 
Note that the sign of the magnetic field determines the sign of the heat current $\ave{J_{\rm H}}=\ave{C_3}-B\ave{J_{\rm S}}$. 
All main features of the B-dependence are semi-quantitatively
reproduced by the truncated GGE with $N_{\rm C}=6$. For very large magnetic fields the convergence to the steady state fixed point becomes slow as transitions rates connecting sectors with different magnetization are strongly suppressed, see Methods for further details.

\vspace{.3cm}
\noindent{\bf \large Discussion}\\
We have demonstrated that driving approximately 
integrable systems activates and pumps into approximately conserved quantities.
Perhaps the most simple experimental setup to measure the pumping effect predicted in this work, is to use a terahertz laser that excites a spin chain material like Cu-benzoate where by symmetry staggered terms of the form (\ref{EqHd}) are expected \cite{affleck99,nojiri06}.
As a consequence of the induced heat currents it is anticipated that the system cools down on one side while it heats up on the other. The direction of the effect can be controlled either by changing the direction of the laser beam or the sign of the external magnetic field $B$. 

For the chosen parameters, the spin and heat currents expressed in dimensionless units appear to be rather small of the order of $10^{-3}$. While these values can definitely be increased  by tuning parameters, for example the external magnetic field, it is important to note that the currents are actually quite large compared to the typical heat or spin currents obtained in bulk materials.
To create a heat current of similar size in a good heat conductor like Cu (assuming $J\sim k_{\rm B} \cdot 100\,$K, $5\,$\AA\ for the distance of the spin chains, and $\kappa^{\rm Cu}\approx 400$\,Wm$^{-1}$K$^{-1}$) one would need a temperature gradient of several $10^5\,$Km$^{-1}$.
Similarly, to create a (transversal) spin current of comparable size in a heavy element like Pt using the spin-Hall effect (assuming $\rho^{\rm Pt}\approx 10$\,$\mu \Omega$\,cm and $\alpha^{\rm Pt}_{\rm s}\approx10$\,\% for the spin Hall angle \cite{sinova15}) one needs electric fields of the order of $10^4\,$Vm$^{-1}$ or sizable current densities of the order of $10^{11}$\,A${\rm m}^{-2}$. These numbers are even more remarkable when one takes into account that the electron densities in Cu or Pt are at least an order of magnitude  higher than the spin density for spin-chains with a distance of $5$\,\AA.

While our study has focused on the steady state, it is instructive to discuss the relevant time scales for its buildup. For this argument we consider a quench where at time $t=0$ an initial state is perturbed both by the integrable part of the Hamiltonian and by small non-integrable perturbations. 
At short times of the order of several $1/J$ the initial state will prethermalize \cite{berges04,moeckel08,kollar11,essler14,bertini15} into a GGE where the values of the conserved quantities, $\langle C_i \rangle$, are set by the initial conditions (with small corrections from the perturbations \cite{essler14,mierzejewski15a}). Further time evolution can be approximately described by a GGE with time-dependent Lagrange parameters. Their time-dependence is determined by perturbations which assert forces $F_i\sim\epsilon^2$, such that $d{\lambda_i}/dt\approx F_i$. Governed by the perturbations the system will loose the memory of its initial condition on a time scale of order $1/\epsilon^2$ and relax to the steady state (obtained from $F_i =0$) which is, in general, completely unrelated to the prethermalized state. Note that the same approach predicts ordinary thermalization in the absence of external driving.

Our results \comm{suggest} that the concept of generalized Gibbs ensembles has a much broader range of application than previously anticipated,
\comm{now extended to open systems where symmetries are not exact and integrability is weakly broken}. 
\comm{A truncated GGE proved to be useful for qualitative description, however, it showed quantitative discrepancies most probably due to  disregarded quasi-local conserved quantities, as observed already in quench protocols \cite{wouters14,pozsgay14}.
We are planning a future study tailored to address this issue systematically.} 
It would be interesting to develop integrability-based methods similar to the quench-action approach \cite{caux13,caux16,wouters14,pozsgay14} to treat such situations.% and relate it to the description of non-equilibrium transport \cite{deluca16}.

Most important for applications is that the integrability is not required to be realized exactly but only approximately. 
Efficient pumping requires only that the pumping rates are of the same order of magnitude as the loss rates arising from integrability breaking terms. Especially the integrability based creation of large spin currents could find its application in future spintronics devices.

\vspace{.7cm}
\noindent{\bf \large Methods}\\
\noindent{\bf Perturbing around $\boldsymbol{\rho_0}$.}
The  central equations (\ref{EqCond1a})  or (\ref{EqCond1b}), used to determine the density matrix $\rho_0$ in the limit $\epsilon \to 0$, have to be consistent and can also even be derived by considering perturbations around
$\rho_0$, $\rho_\infty=\rho_0+ \delta \rho$.

First, the leading $\delta\rho$ correction to $\langle \dot C_i \rangle$, Eq.~(\ref{EqCond1}), arising from $\tr(C_i  \cl{L}_0  \delta \rho)$ which is nominally of the same order as $\tr(C_i  \epsilon \cl{L}_1 \rho_0)$ vanishes trivially as $\tr(C_i [H_0, \delta \rho])=\tr(\delta \rho [C_i,H_0])=0$. 

For arbitrary $\rho_0$, $\delta \rho$ is exactly given by $\delta\rho=-\cl{L}^{-1}\epsilon \cl{L}_{1}\rho_0$, where $\cl{L}^{-1}$ is a short-hand notation for 
$\lim_{\eta\to 0}(\cl{L}-\eta \hat{1})^{-1}$ with the infinitesimal regularizer $\eta$. 
The correct expansion point $\rho_0$ is found if $\lim_{\epsilon \to 0} \delta \rho=0$. Below we show that for the projection operator $\hat P$, Eq.~(\ref{EqP}),
\begin{equation}\label{EqLP0}
\cl{L}^{-1}\hat{P}\sim \mathcal{O}(\epsilon^{-1}),	
\end{equation}
which would yield 
$\cl{L}^{-1} \hat{P} \epsilon \cl{L}_1 \rho_0 \sim \mathcal{O}(1)$. This contradicts our perturbative approach
unless $\hat{P} \cl{L}_1 \rho_0=0$, as set by our condition Eq.~(\ref{EqCond1a}).

Eq.~(\ref{EqLP0}) is a consequence of the fact that $\hat P$ projects onto the tangential space to GGE density matrix. In this space $\cl{L}_0$ vanishes by definition, $\cl{L}_0 (\partial\rho_0/\partial \lambda_i)=0$, and
$\cl{L}=\cl{L}_0+\epsilon \cl{L}_1$ is  therefore of order $\epsilon$. Technically, this can be seen by using the general relation
\begin{equation}
(\hat X + \hat Y)^{-1} - \hat X^{-1} 
%=(X+Y)^{-1} (X- (X+Y))X^{-1} 
= - (\hat X + \hat Y)^{-1} \hat Y \hat X^{-1} 	
\end{equation}
for 
\begin{align}
&\hat{X}=\hat{P}\epsilon \cl{L}_1 \hat{P} ,\notag \\
&\hat{Y}=\cl{L}_0+\hat{Q}\epsilon \cl{L}_1 \hat{Q}
+\hat{P}\epsilon \cl{L}_1\hat{Q}+\hat{Q}\epsilon \cl{L}_1\hat{P},
\end{align}
with $\hat Q=\hat 1-\hat P$ and $\hat{X}+\hat{Y}=\cl{L}$. Then
\begin{align}
\cl{L}^{-1} \hat{P} 
&=(\hat{X}+\hat{Y})^{-1} \hat{P} \notag \\
%&=\hat{X}^{-1} \hat{P} - (\hat{X}+\hat{Y})^{-1} \hat{Y} \hat{X}^{-1} \hat{P}\notag \\
&=\hat{X}^{-1} \hat{P} - (\hat{X}+\hat{Y})^{-1} \ \hat{Q} \ \hat{Y} \ \hat{P} \ \hat{X}^{-1} \hat{P}\notag \\
&\sim \mathcal{O}(\epsilon^{-1}) + \mathcal{O}(1)	
\end{align}
The second term is $O(1)$ as $\cl{L}_0 \hat P=0$ and therefore $\hat Y \hat P \sim O(\epsilon)$. The divergence of $\cl{L}^{-1} \hat P$ for $\epsilon \to 0$ can be directly related to the fact that integrable systems are characterized by infinite conductivities (finite Drude weights) at finite temperatures \cite{zotos99} as can, e.g., be seen \cite{jung06} within the memory matrix formalism \cite{forster75}.

All arguments given above can be generalized to situations where leading corrections arise from 2nd order perturbation theory in which case one obtains Eq.~(\ref{EqCond1b}) instead of Eq.~(\ref{EqCond1a}).

\vspace{.3cm}
\noindent{\bf Staggered hopping and magnetic field modulation.}
Sizable staggered g-tensors leading to staggered B-fields have been observed in a number of different compounds \cite{affleck99,nojiri06,kimura13,niesen14}. Similarly an external electric field will distort the crystalline structure in these materials, leading to staggered exchange couplings linear in homogeneous electric fields. An example of such a material is Cu-benzoate \cite{affleck99} with the above modulations allowed by symmetry for electric (magnetic) fields applied in the 010 (001) crystallographic direction. In this system the staggered g-tensor has been measured to be approximately 0.08 \cite{nojiri06}, the size of the staggered exchange coupling is unknown. For simplicity, we assume in Eq.~(\ref{EqHd}) that the two staggered terms are of the same size.

\vspace{.3cm}
\noindent{\bf Conservation laws of the XXZ Heisenberg model.} 
An infinite set of local conserved quantities $C_i$ of the Heisenberg model 
$H_{\rm XXZ}=H_0(B=0)$
can be obtained using the boost operator $O_{\rm b}=-i\sum_{j} j h_{j,j+1}$ (where $H_{\rm XXZ}=\sum_j h_{j,j+1}$) from the recursion relation 
$[O_{\rm b},C_i]= C_{i+1}$ for $i>1$ with $C_1=\sum_j S^z_j$, $C_{2}=H_{\text{XXZ}}$ \cite{grabowski95}. In general, $C_i$ are operators involving maximally $i$ neighboring sites. Importantly, $C_3$ in the absence of external magnetic field equals the heat current

\begin{align}\label{EqJhs}
&J_{\rm H}(B=0)=C_3=J^2 \sum_j(\boldsymbol S'_j \times \boldsymbol S''_{j+1})\cdot\boldsymbol S'_{j+2}	
\end{align}
with rescaled spin operators $S'^{a}_j=\sqrt{\lambda_a} S^a_j,
S''^{a}_j=\sqrt{\lambda_z/\lambda_a} S^a_j$ for $
\lambda_z=\Delta/J$, $\lambda_x=\lambda_y=1.$ 
In the presence of external magnetic field, Eq.~(\ref{EqXXZ}), heat current has in addition to $C_3$ also a spin current component,
\begin{equation}
J_{\rm H}=J^2 \sum_j(\boldsymbol S'_j \times \boldsymbol S''_{j+1})\cdot\boldsymbol S'_{j+2} - B J_{\rm S}. 
\end{equation}

As understood recently there also exist families of quasi-local conserved quantities \cite{prosen13,mierzejewski15,ilievski15}, which are mostly disregarded in our study with the exception of a spin-reversal parity-odd operator, $J_{\rm S}^{\rm c}$. The latter is constructed as the conserved part of the spin current operator $J_{\rm S}$,
\begin{align}\label{EqJs}
J_{\rm S}&=i \frac{J}{2} \sum_j(S^+_j S^-_{j+1} - S^-_j S^+_{j+1}) \\
J_{\rm S}^c&= \sum_{\tilde n} |\tilde  n\rangle \langle \tilde n|J_S|\tilde n \rangle \langle \tilde n| \notag
\end{align}
where $|\tilde n\rangle$ are simultaneous eigenstates of the $ C_i$. 
Since it is known that the spin current has an overlap with the quasi-local family \cite{prosen11} for $\Delta<J$, the conserved $$J_\text{S}^\text{c}$$ contains quasi-local components (and, possibly, non-local components not contributing in the thermodynamic limit). 

\vspace{.3cm}
\noindent{\bf Floquet formulation.} 
For a periodically driven system described by $\dot \rho=\cl{L}(t) \rho$ with $\cl{L}(t+T)=\cl{L}(t)$ 
the density matrix changes periodically in the long-time limit. Therefore it is useful to split it into Floquet components,

\begin{align}\label{EqFloquetIndex}
\rho=\sum_{n} e^{-in\omega t} \rho^{(n)}, n \in \mathbb{Z}
\end{align} 
with $\rho^{(-n)}={\rho^{(n)}}^\dagger$ and $\omega=2 \pi/T$. The Floquet components are combined into  the vector $\boldsymbol \rho=(\dots \rho^{(-1)},\rho^{(0)}, \rho^{(1)},\dots) $. The Liouvillian is promoted to a (static) matrix $\boldsymbol{\cl{L}}_{nm}=i n \omega \delta_{nm}+\cl{L}_{n-m}$ with $\cl{L}_{n-m}=\frac{1}{T} \int_0^T \cl{L}(t) e^{i \omega (n-m) t} d t$.
Using this notation, all results obtained for static Liouvillian super-operators directly 
translate to the time-periodic case. Within our setup, $H_0$, all approximate conservation laws $C_i$ and the GGE density matrix $\rho_0$ are static and therefore the projection operator $\hat P$, Eq.~(\ref{EqP}), projects onto the $n=0$ Floquet sector only. The steady state condition,  Eq.~(\ref{EqCond1b}), thus means that
the approximately conserved quantities do not grow after averaging over an oscillation period. To second order in $\epsilon_{\rm d}$ only transitions from the $n=0$ to the $n=\pm 1$ Floquet sector and back contribute to   Eq.~(\ref{EqCond1b})  or (\ref{EqLambdadynamics}) as $\cl{L}_{n}=0$ for $|n|>1$. 

For the generalized force due to the periodic driving, we obtain from (\ref{EqLambdadynamics})
\begin{align}\label{FdLehmann}
F_{i}^{\rm (d)}&=
\frac{2\pi}{N} \epsilon^2_{\rm d} \ \sum_{i'}(\chi^{-1})_{ii'} 
\sum_{m,k} \rho_m(C_{i',m}-C_{i',k}) \ \times \notag\\
\times \ \Big\{
&|\langle k|H_{\rm d}^{(+)}|m\rangle|^2 \delta(E^0_k-E^0_m-\omega) \notag \\
+&|\langle k|H_{\rm d}^{(-)}|m\rangle|^2 \delta(E^0_k-E^0_m+\omega)
\Big\} 
\end{align}
where we used $H_0$ eigenstates $|m\rangle$ with $H_0 |m\rangle=E_m^0|m\rangle$, matrix elements $\rho_m=\ave{m|\rho_0|m}$, $C_{i,m}=\ave{m|C_i|m}$, and the notation $H_{\rm d}=\epsilon_{\rm d} \left(e^{i \omega t} H_{\rm d}^{(-)} + e^{-i \omega t} H_{\rm d}^{(+)}\right)$.  Note that Eq.~(\ref{FdLehmann}) contains -- as expected --  transition rates well-known from Fermi's golden rule. Eq.~(\ref{FdLehmann}) is evaluated for finite systems of size $N$  by replacing the $\delta$ function by a Lorentzian $(1/\pi) \eta/(\omega^2+{\eta}^2)$ ($\eta=0.1J$ for  $N=12$). 

Eq.~(\ref{FdLehmann}) is only valid for situations where all conservation laws commute with each other, with $C_i=\sum_m |m\rangle C_{i,m} \langle m |$, see below for a brief discussion of the non-commuting case.

\vspace{.3cm}
\noindent{\bf Phonon coupling.} As written in the main text, we assume that the phonon system always remains at equilibrium,  $\rho_{\rm ph} \sim e^{-H_0^{\rm ph}/T_{\rm ph}}$.
%The total density matrix is therefore approximately a tensor product of $\rho_{\rm ph}$ and the density matrix $\rho$ of the spin system,  $\rho_T= \rho \otimes \rho_{\rm ph}$.
Using Eq.~(\ref{EqLambdadynamics}), after tracing over phonons, we obtain for the generalized force  
\begin{align}\label{EqFph}
&F_i^{\rm (ph)}=
2\pi \epsilon^2_{\rm ph} \ \sum_{i'}(\chi^{-1})_{ii'} 
\sum_{m,k} \rho_m(C_{i',m}-C_{i',k})   \notag \\
 &\times J^2\left(|\langle k|\mathbf{S}_j \cdot \mathbf{S}_{j+1}|m\rangle|^2
   +\gamma_{\rm m}^2|\langle k|S^x_j S^z_{j+1} + S^z_j S^x_{j+1}|m\rangle|^2\right) \notag \\
&\times 
\big((n_{\rm B}(E^0_m-E^0_k)+1) \ A^{\rm (ph)}(E^0_m-E^0_k) \notag \\ 
& \qquad  + n_{\rm B}(E^0_k-E^0_m) \ A^{\rm (ph)}(E^0_k-E^0_m)\big)
\end{align}
where $n_{\rm B}(E)=1/(e^{E/T_{\rm ph}}-1)$ is the equilibrium Bose distribution evaluated at the temperature $T_{\rm ph}$ and $A^{\rm (ph)}(\omega)$ is the phonon spectral function. For our finite size calculation we broaden the spectral function of the Einstein phonons using $A^{({\rm ph})}(\omega)=\Theta(\omega) \frac{\omega}{\omega_{\rm ph} \eta \sqrt{\pi}} e^{-(\omega-\omega_{\rm ph})^2/\eta^2}$. This choice of broadening ensures detailed balance relations (necessary to obtain a thermal state in the absence of driving)
and the positivity of phonon frequencies (necessary for stability).
{
%Fig.~\ref{fig3} and Fig.~\ref{fig4} are calculated with broadening $\eta=0.1$, while Fig.~\ref{fig5} is obtained with $\eta=0.4$ 
For all plots we use $\eta=0.4J$. However, we have checked that similar results are obtained, e.g., for $\eta=0.1J$ for magnetic fields up to $|B|=2J$. For larger fields $\eta=0.1J$ does not provide a sufficient amount of relaxation between sectors with different magnetization and convergence becomes slow and unstable. For $\eta=0.4J$ larger fields, $|B| \lesssim 5J$, can be reached.

\vspace{.3cm}
\noindent{\bf Implementation of non-commuting conservation laws.}
As discussed in the main text, a complete basis of all non-local commuting or non-commuting conserved quantities is given by $\mathcal Q=\{|n\rangle \langle m| \ \text{with}\  E_m^0=E_n^0\}$ which solve the equation $\cl{L}_0 Q_i=0$ for $Q_i \in \mathcal Q$. Using the exact eigenstates of $H_0$ it is straightforward to evaluate Eq.~(\ref{EqRhoExact}) where we use for our finite size calculations the broadening procedures described above. As a technical detail we note that, when one follows this procedure, one has to evaluate in the phonon sectors integrals of the type $\int \frac{A^{(\rm ph)}(\omega')}{\omega-\omega'} n_{\rm B}(\omega') d\omega'$ numerically. For efficient evaluations we use interpolating functions for these integrals.

\vspace{.3cm}
\noindent{\bf GGE estimation for other conserved quantities.}
To provide further support for our claim that truncated GGEs give a semi-quantitative description of our weakly open system we show in Fig.~\ref{fig6} additional comparison of the $\ave{H_0}$ and $\ave{C_4}$ as a function of magnetic field $B$ at $(\epsilon_{\rm d}/\epsilon_{\rm ph})^2=2.5$, comparing as in the main text the exact calculation including all conserved quantities and the truncated GGE with $N_{\rm C}=6$ (quasi-)local conserved quantities. The GGE ansatz captures the right magnitude and the correct behaviour in the dependence on $B$ also for more complicated 4-spin operators like $C_4$. We use same parameters as for the Fig.~\ref{fig5} in the main text: $(\epsilon_{\rm d}/\epsilon_{\rm ph})^2=2.5, J=1,\Delta=0.8,\omega=1.6 \ \omega_{\rm ph}, \omega_{\rm ph}=T_{\rm ph}=1, N=12$.

\begin{figure}[!t]
\center \includegraphics[width=.92\linewidth]{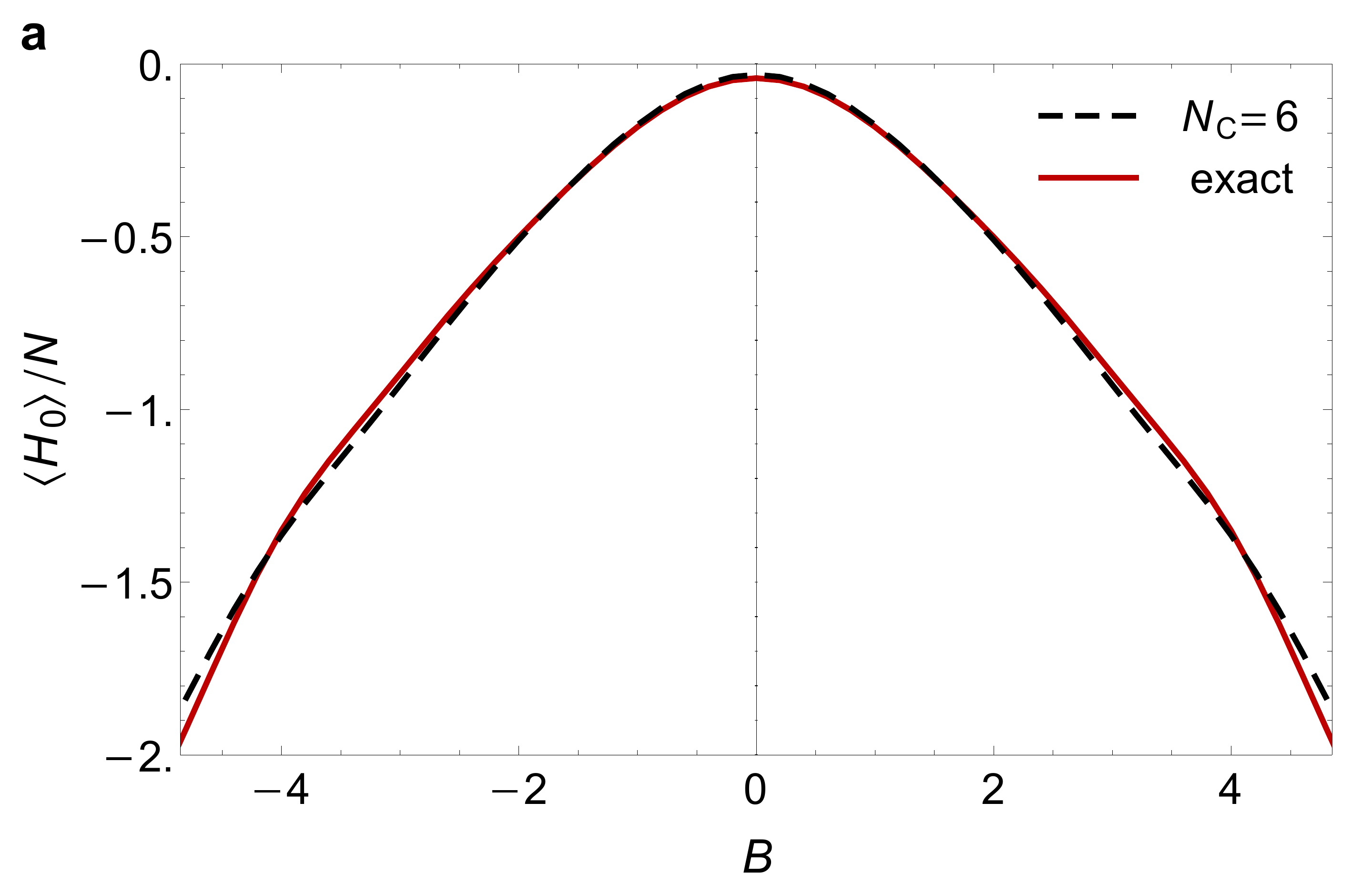}
\includegraphics[width=.92\linewidth]{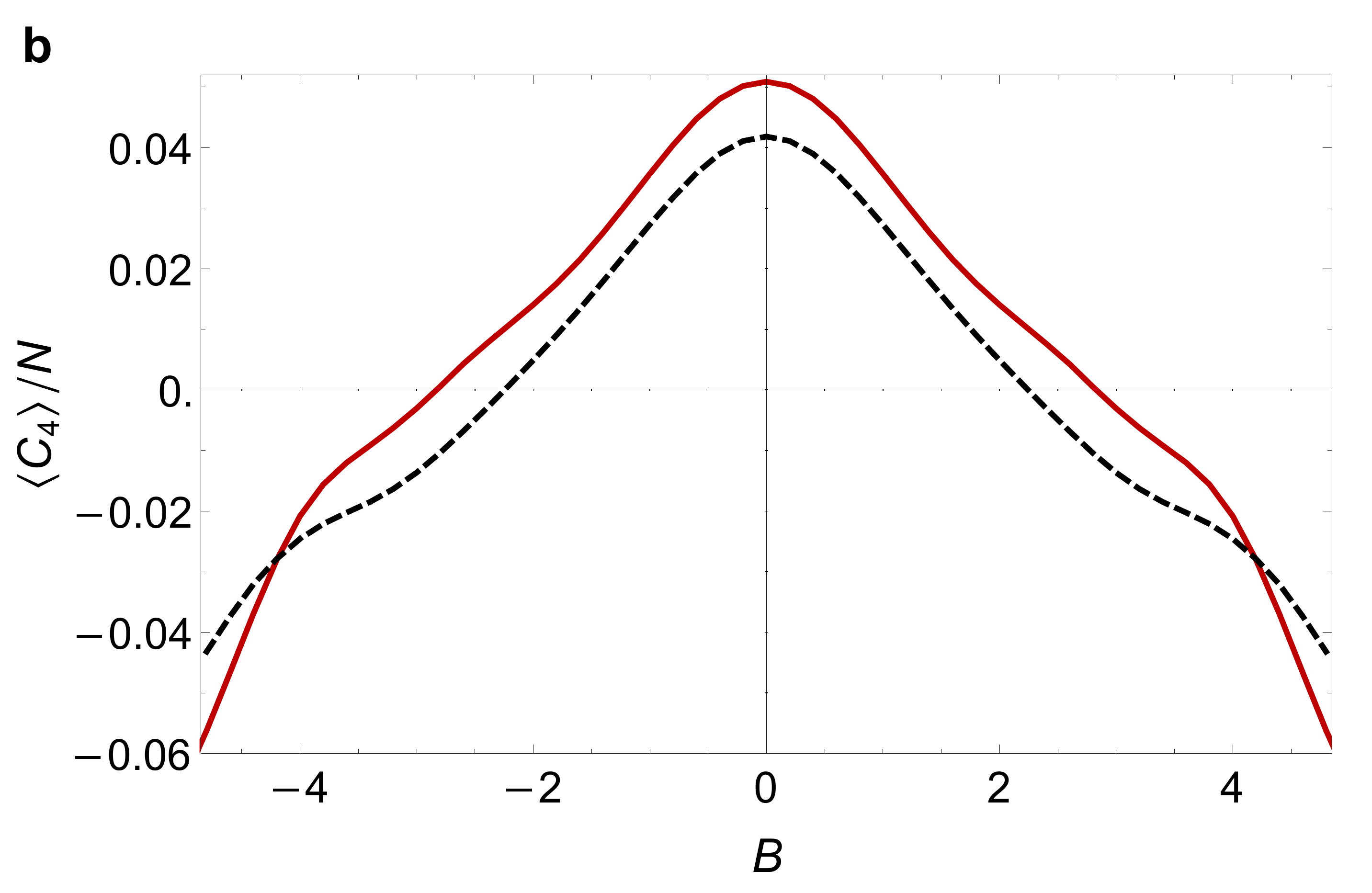}
\caption{(a) The energy density and (b) the expectation value of  another conserved quantity $C_4$ (4-spin operator) as a function of magnetic field $B$, obtained from calculation using all conserved quantities (solid) and a GGE with $N_{\rm C}=6$ (quasi-) local conserved quantities (dashed).
\label{fig6}}
\end{figure}

\vspace{.3cm}
\noindent{\bf \large Acknowledgements}\\ \noindent
We acknowledge useful discussions with S. Diehl, F. H. L. Essler, M. Fagotti,  E. Ilievski, M. Mierzejewski, J. De Nardis, T. Prosen, and M. C. Rudner, H. F. Legg for reading the manuscript, and financial support of the German Science Foundation under
CRC 1238 (project C04) and CRC TR 183 (project A01).
\\

\vspace{.15cm}
\noindent{\bf \large  Author contributions}\\ \noindent
A. R. and Z. L. designed the study, Z. L. and F. L. performed analytical calculations and 
F. L. implemented the numerical codes, all authors analyzed the results and contributed to the manuscript.

%\vspace{.2cm}
%\noindent{\bf \large  Additional information}\\ \noindent
%{\small
%{\bf Competing financial interests:} The authors declare no competing financial interests.

%\vspace{.1cm}
%\noindent{\bf Code availability:} Even though we used only standard numerical routines to produce our numerical results, some further guidance can be provided.
}
\end{document}